# Spin-flip-driven giant magneto-transport in A-type antiferromagnet NaCrTe$_2$

**Junjie Wang,**[1,2,+] **Jun Deng,**[1,2,+] **Xiaowei Liang,**[3] **Guoying Gao,**[3*] **Tianping Ying,**[4] **Shangjie Tian,**[5] **Hechang Lei,**[5] **Yanpeng Song,**[1] **Xu Chen,**[1] **Jian-gang Guo**[1,6*] **and Xiaolong Chen**[1,2,6*]

[1] Beijing National Laboratory for Condensed Matter Physics, Institute of Physics, Chinese Academy of Sciences, Beijing 100190, China

[2] University of Chinese Academy of Sciences, Beijing 100049, China

[3] Center for High Pressure Science, State Key Laboratory of Metastable Materials Science and Technology, Yanshan University, Qinhuangdao 066004, China

[4] Materials Research Center for Element Strategy, Tokyo Institute of Technology, Yokohama 226-8503, Japan

[5] Department of Physics and Beijing Key Laboratory of Opto-electronic Functional Materials & Micro-nano Devices, Renmin University of China, Beijing 100872, China

[6] Songshan Lake Materials Laboratory, Dongguan, Guangdong 523808, China

## ABSTRACT

**For anisotropic magneto-resistance (AMR) effect, its value synergistically depends on the magnitudes of magneto-resistance (MR) and magneto-crystalline anisotropy energy (MAE) simultaneously. In a magnetic material, the concurrence of gigantic AMR and MR signals is rather difficult due to weak spin-lattice coupling and small MAE. Here we report the considerable magneto-transport effect in layered A-type antiferromagnetic (AFM) NaCrTe$_2$ by realigning the spin configurations. By applying *H*, the antiparallel spins of adjacent layers are flipped to ferromagnetic (FM) coupling either Ising-type along *c*-axis or XY-type within *ab*-plane. Theoretical calculations reveal that the energy bandgap narrows from 0.39 eV to 0.11 eV, accompanying a transition from semiconductor (high-*R* state) and half-semiconductor (low-*R* state), respectively. Thus, gigantic negative MR ratio of -90% is obtained at 10 K. More importantly, the decrement of *R* along *H*//*c* is far quicker than that of *H*//*ab* because the MAE of Ising-FM state is 1017 μeV/Cr$^{3+}$ lower than that of XY-FM. The distinct trends result in the AMR ratio of 732% at 10 K, which is the record value to our best knowledge. These findings**

unravel the intrinsic origin of magneto in NaCrTe$_2$ and will stimulate us to exploring the *H*-sensitive transport property in more AFM materials.

## I. Introduction

Anisotropic magneto-resistance (AMR) origins from the relativistic spin-orbit coupling (SOC) in magnetic materials, which is inherently determined by the variation of conductivity and density of states at the Fermi level ($E_F$) under *H* [1]. Normally, the resistance (*R*) is dependent on the relative angle between the electric current (*J*) or crystallographic axes and *H*, where the *R* under parallel *H* is usually larger than that of perpendicular direction in a ferromagnetic alloy. The AMR magnitudes of a few percent (<5%) are found in known ferromagnetic Permalloy [2]. For practical applications, this kind of spin-dependent transport property has enormous technological importance, particularly in fabricating magnetic memory devices for recording [3] and sensor [4].

Exploring strong AMR effect in antiferromagnetic (AFM) materials have been attracting intensive interests due to their superior properties like flipped spin, reduced stray field and fast spin dynamics [5,6]. However, altering the AFM ordering and controlling the *R* in antiferromagnets by *H* are generally difficult, usually leading to weak AMR signal. This has been reported in a series of collinear AFM alloys and compounds like FeRh [7], CuMnAs [8] and MnTe [9]. Few AFM materials with large AMR values are identified. In AFM oxides La$_{2-x}$Sr$_x$CuO$_4$ [10] and Pr$_{1.3-x}$La$_{0.7}$Ce$_x$CuO$_4$ [11], strong coupling between spin and lattice modifies the electronic structure as in-plane spin-flop transitions happen. This kind of *H*-direction dependent conductivity leads to an AMR ratio of 28%. Recently, another $J = 1/2$ AFM layered semiconductor Sr$_2$IrO$_4$ is intensively investigated, in which the AMR signal is observed in the regime of in-plane meta-magnetic transition [12],[13],[14],[15]. In the latest report, the largest AMR ratio of 160% is achieved in high-quality Sr$_2$IrO$_4$ single crystal [16]. Theoretical calculations reveal that as the net spins of Ir$^{4+}$ change from [110] to [100] axis, the energy bandgap ($E_g$) of 35.0 meV decreases to 25.9 meV. The *H*-dependent electronic structure is also the main origin for AMR. Based on above analyses, it is suggested that the easy-flipped magnetic order in AFM compounds are decisive factors for in turn tuning conductivity and inducing large AMR effect.

As for different magnetic orders in antiferromagnets, the A-type AFM state is the simplest one with intralayer FM and interlayer AFM coupling, as shown in Fig. 1a. There are two spin-flip fields, $H_{sf1}$ and $H_{sf2}$, along magnetic easy-axis (out-of-plane) and hard-axis (in-plane), respectively. Under $H > H_{sf1}$, the A-type AFM coupling will change into an Ising-like FM coupling, where all the spins of the atoms point to the *c*-axis. While for $H > H_{sf2}$, the spin directions lie in the *ab*-plane, leading to a XY-like FM coupling. In the terms of energy E, the A-type AFM has the lowest energy. Both FM alignments are excited states, in which the energy of Ising-type FM state is lower than that of XY-type FM, as depicted in Fig. 1b. This is like the effect of giant magneto-resistance (GMR), where the change of magnetic coupling leads to different resistances.

Following this anticipation, we synthesize the single crystal of $NaCrTe_2$ with layered A-type AFM order and investigate its low-temperature magneto-transport properties. As *H* exceeds the spin-flip fields $H_{sf}$ of easy-axis (3.2 T) and hard-axis (13.0 T), the antiparallel spins of adjacent layers realign to Ising-like FM and XY-like FM, respectively. Meanwhile, the calculated $E_g$ reduces from 0.39 eV (high-*R* state) to 0.11 eV (low-*R* state). The enhanced conductivity results in a negative MR of -90%. Furthermore, owing to the large MAE between two FM states, the descending trends from high-*R* to low-*R* are far different, resulting in the record AMR ratio of 732%. Our works highlight the importance of narrow-bandgap antiferromagnet for exploring *H*-sensitive magneto-property.

## II. Experiment

Stoichiometric amount of Na, Cr, and Te were mixed and loaded into an alumina crucible that was sealed into evacuated quartz tube with back-filled 0.2 atm of argon. The assembly was heated to 1000 K with duration of 25 h. After that, the sintered sample was taken out, reground, pelletized and re-heated to 1000 K for 25 h for homogeneity. For growing $NaCrTe_2$ single crystal, the sealed quartz tube was heated to 1200 K with duration of 20 h, slowly cooled to 1000 K with rate of 3 K/h. Then it was finally furnace cooling to room temperature. Single crystal with shiny surface and size of 1.5 ×1.5 ×0.1 $mm^3$ was obtained. All the procedures are handled in a glove-box filled with argon gas.

Powder X-ray diffraction (PXRD) patterns were collected using a PANalytical X'Pert PRO diffractometer (Cu Kα-radiation) with a graphite monochromator in a reflection mode. Rietveld refinement of PXRD pattern was performed using *Fullprof* software suites [17]. The scanning electron microscopy (SEM) image of the sample was captured from Hitachi S-4800 FE-SEM. The composition was determined by Energy Dispersive Spectroscopy (EDS) with average of 10 sets data. Electrical resistivity (*ρ*) and Hall resistivity ($ρ_{xy}$) were measured through standard six-wire method using the physical property measurement system (PPMS-9T and -14T, Quantum Design), respectively. Magnetic properties were measured through vibrating sample magnetometer (VSM) option in the PPMS-9T and PPMS-14T, Quantum Design.

First principles calculations were performed with the Vienna *ab*-initio simulation package (VASP) [18]. We adopted the generalized gradient approximation (GGA) in the form of the Perdew-Burke-Ernzerhof (PBE) [19] and HSE06 [20] for the exchange-correlation potentials. A Hubbard $U = 3$ eV was adopted for Cr-3$d$ orbital. The projector augmented-wave (PAW) pseudopotentials were used with a plane wave energy 500 eV. $2p^63s^1$, $3d^54s^1$ and $5s^25p^4$ electron configurations were treated as valence electrons for Na, Cr and Te, respectively. A Monkhost-Pack Brillouin zone sampling grid with a resolution of $0.02\times2\pi$ Å$^{-1}$ was applied. Parameters were relaxed until all the forces on the ions were less than $10^{-3}$ eV/Å. Phonon spectra were calculated using finite displacement method implemented in the PHONOPY code [21] to determine the dynamical stability of the studied structures. The magnetism is described by the classical Heisenberg Hamiltonian $H = -J_1 \sum_{i,j} S_i S_j - J_2 \sum_{i,j} S_i S_j - J_3 \sum_{i,j} S_i S_j$, where $J_1$ is the nearest neighbor, $J_2$ the next-nearest neighbor and $J_3$ interlayer coupling constant, respectively. These constants can be extracted from the total energy of different magnetic structures.

III. Results and discussion

A. Spin-flip transitions in NaCrTe$_2$

Fig. 1c shows crystal structure of A-type AFM NaCrTe$_2$, in which the CrTe$_6$ octahedrons are edge-shared in CrTe$_2$ layer and the Na ions locate between CrTe$_2$ layers. The Rietveld refinements of PXRD pattern of NaCrTe$_2$ is shown in Fig. S1a. NaCrTe$_2$ crystallizes in the space group of *P*-3*m*1 with $a = 4.0053$ Å and $c = 7.4485$ Å. Based on the formation enthalpies and phonon spectra shown in Fig. S1b-1d, the composition and structure of NaCrTe$_2$ are thermodynamically stable. We calculate the total energy of both FM states and other three AFM states by using 2×√3 supercell (Fig. S2). For the A-type AFM, the extracted magnetic exchange constants of the $J_1$, $J_2$ and $J_3$ are 2.98 meV, 1.86 meV and -0.08 meV, see Table 1. These values indicate that the intralayer and interlayer magnetic coupling are FM and AFM, respectively, and the interlayer interaction is very weak. Furthermore, it is found that the A-type AFM has the lowest energy of -140.8182 eV. The Ising-FM and XY-FM coupling are 5.9 meV and 19.8 meV higher in energy, respectively.

High-quality NaCrTe$_2$ single crystal is grown by self-flux method, and its PXRD pattern and optical image are shown in Fig. S3a. The EDS spectra and chemical-element mapping confirm that the atomic ratio is close to 1:1:2 as shown in Fig. S3b-3f. The temperature-dependent magnetic susceptibility ($\chi$) of NaCrTe$_2$ single crystal under $H//c$ and $H//ab$ are plotted in Fig. 1d. At $H$ = 1 T, the features of $\chi$-T curves indicate that intralayer and interlayer interactions are FM and AFM couplings, respectively, and the Néel temperature $T_N$ is 106 K. It matches the above predication and other report [22]. In the Fig. S4, the effective moment $\mu_{eff}$ per Cr$^{3+}$ obtained from Curie-Weiss fitting is 3.91 $\mu_B$ for $H // c$, which is consistent with the spin-only moment (3.88 $\mu_B$) in a free Cr$^{3+}$ ion. The fitting equations are shown in Supplementary Information (SI) section S1. As increasing $H$ along $c$-axis, the $T_N$ gradually lowers. As $H$ > 3 T, the AFM transition totally disappears, and then an $H$-induced FM transition emerges. The $\chi$ in $ab$-plane, however, is hardly suppressed as increasing $H$ to 5 T. The M-H curves in Fig. 1e and 1f exhibit paramagnetic (PM) behaviors at $T$ > $T_N$, while the M abruptly jumps to saturated values in the range of 2 T< $H$ < 3 T below $T_N$ with $H // c$, indicating a transition from A-type AFM to Ising-type FM along the $c$-axis. At $H$ = 5 T, the saturated FM moment is 3.27 $\mu_B$ at 10 K. For $H // ab$, it also exhibits spin-flip transition from A-type AFM to XY-type FM below $T_N$, and the moment slowly increases and then saturates to 1.64 $\mu_B$ at 10 K as $H$ >8 T.

**B. MR and AMR effects**

Fig. 2a and 2b show the temperature-dependent in-plane $\rho_{ab}$ and out-of-plane $\rho_c$ under $H // c$. At low-$H$, there are large jumps in $\rho_{ab}$ and $\rho_c$ curves at $T_N$, which are associated with MIT-like transition due to the formation of A-type AFM ordering. As increasing $H$, the $T_N$ gradually shifts to low temperatures and the amplitude of resistivity jump decreases. At $H$ > 3 T, this resistivity jump is suppressed, and then the small kinks in $\rho_{ab}$ and $\rho_c$ show up. It might be viewed as manifestations of $H$-induced FM transition. The derivation of $\rho_{ab}$ and $\rho_c$ of $H$ = 5 T are shown in Fig. S5, in which the peaks due to FM-like coupling locate at $T_c'$ = 113-122 K. In Fig. 2c and 2d, we show the isothermal in-plane MR of NaCrTe$_2$ over a broad $T$ (10 K- 300 K) and $H$ (up to 14 T) along magnetic $c$-axis and $ab$-plane. Here we define the MR ratio as

$$MR = \frac{\rho(H) - \rho(0)}{\rho(0)} \times 100\%$$

We find that, at $T$ > $T_N$, the absolute value of MR is generally small. Below $T$ < $T_N$, a kind of step-like decrement in MR happens as $H$ > $H_{sf1}$, resulting in negative MR effects along both directions. The maximal MR$_{ab}$ are -81.6 % and -78.5% at 10 K under $H // c$ and $H // ab$. The maximal MR$_c$ are as large as -87.9% and -90.6% for $H // c$ and $H // ab$ as shown in Fig. S6a and S6b, respectively. Such MR is intimately linked to the different conductivity between AFM and FM coupling of NaCrTe$_2$. Therefore, we measured the $H$-dependent Hall resistivity $\rho_{xy}$ from 10 K to 300 K and calculated carrier concentration $n$ based on a single-band model. The data are plotted in Fig. S6c and S6d. The fitting equations are shown in SI section S2. Under PM state, the $H$-dependent $\rho_{xy}$ is linear. Under low-$H$, the slope abruptly increases below $T_N$, suggesting that the $n$ of AFM state reduces. Under high-$H$, the slope of low-T curves inversely lowers, indicating that the $n$ increases after transiting into FM phase. The magnitudes of $n$ increase from $7\times10^{19}$ cm$^{-3}$ to $1.3 \times10^{20}$ cm$^{-3}$, which correspond to the high-$R$ of AFM state and low-R of FM state, respectively.

On the other hand, we find that the decrement of $R$ along $H \mathbin{/\mkern-6mu/} c$ is remarkably quicker than that of $H \mathbin{/\mkern-6mu/} ab$. It implies that the A-type AFM is easier flipped to Ising-type FM than XY-type FM, evidenced by the smaller $H_{sf1}$. In Fig. 2e and 2f, two magnetic phase diagrams are plotted based on the data taken from the MR and $\chi$-T curves. We can see that the A-type AFM firstly transits to an intermediate phase and then changes into Ising-type FM or XY-type FM as increasing $H$. Along $H \mathbin{/\mkern-6mu/} c$, there is small gap of 0.5 T between $H_{sf1}$ and $H_{sf1}$'. While the gap between $H_{sf2}$ and $H_{sf2}$' is 12 T for $H \mathbin{/\mkern-6mu/} ab$. It clearly explains the anisotropy in the process of flipping the A-type AFM state. In the intermediate phase, there may exist mixed spin structure or canted FM phase like the observation in CrPS$_4$ [23].

To check the AMR in NaCrTe$_2$, we systematically measured the angle-dependent MR curves and plotted the data of selected temperatures in Fig. 3. Supplementary AMR data can be seen in the Fig. S7 and S8. Here the AMR ratio is defined as

$$AMR = \frac{R_{max} - R_{min}}{R_{min}} \times 100\%$$

at each temperature, where $R_{max}$ and $R_{min}$ are the maximal and minimal values for a given field as $\theta$ sweeping from 0° to 360°. The measured schemes are drawn in the inset of Fig. 3c and 3d. At $T>T_N$, such as 130 K, the NaCrTe$_2$ is under PM state, and the $R_{ab}$ shows a weakly angle-dependent variation as shown in Fig. S7a. The resultant AMR of 5 T is 6.3%, which is comparable to those values in many FM materials [24] and conventional AFM semiconductor [25]. At $T = 90$ K, just below $T_N$, the $R_{ab}$ of $H = 1$ T initially increases and then decreases, exhibiting two-fold symmetry with maximal values at $\theta = 90°$ and 270° and the minimal values at $\theta = 0°$ and 180°. At $H = 2.7$ T and $\theta = 0°$, the magnetic coupling is A-type AFM and thus the $R_{ab}$ is at high-R state. Increasing $\theta$ to the flipped angle $\theta_f = 30°$, the coupling changes into Ising-FM state, so the $R_{ab}$ slowly lowers and then keeps the low-$R$ state from 45° to 135°. At larger $\theta$, the $R_{ab}$ recovers and repeats the oscillations with two-fold symmetry plus $R$-plateau. Meanwhile, the $R$-oscillation exhibits the largest amplitude at $H = 2.7$ T. Under $H = 9$ T, the two-fold symmetry still survives while the oscillated magnitude becomes weaker again. These features are more remarkable at $T = 10$ K, see right panel of Fig. 3a. All the curves exhibit two-fold symmetry while having wider plateaus because the $\theta_f$ moves to 60°. It means that at lower temperatures, one need stronger $H$-component along $c$-axis to align all the spins to Ising-type FM form. In Fig. 3b, the periodic oscillation of $R_c$ are almost similar to those of $R_{ab}$ except the amplitude. We summarize the $H$-dependent AMR ratios of $R_{ab}$ and $R_c$ in Fig. 3c and 3d. For $R_{ab}$ and $R_c$, the AMR signal are pretty small below $H_{sf1}$, and then quickly reach peaks of 275% and 732% at 2.7 T and 3.2 T for 90 K and 10 K, respectively. As for $R_c$, the maximal AMR ratios of 90 K and 10 K are 170% and 580% at $H = 2.7$ T and $H = 3.2$ T, respectively. At $H = 9$ T, the AMR ratios monotonously decrease to ~20% and ~150% at 90 K and 10 K. Further increasing $H$ to 14 T, the oscillation of $R$ disappear as shown in Fig. S9, thus the AMR ratio becomes zero. Here it is noted that the maximal AMR ratio in our samples are a few times larger than those in $La_{0.7}Ca_{0.3}MnO_3$ (80%) and $Sr_2IrO_4$ (160%) [16], which is the largest value to our best knowledge.

**C. Origin of the large MR and AMR**

To gain a deeper understanding of the origin of large AMR of $NaCrTe_2$, we calculated the electronic band structures and projected density of states (PDOS) of A-type AFM and both FM states by density functional theory with HSE06+U. Spin-orbit-coupling is considered because of heavy element Te. In Fig. 4a, one can see that the band dispersion around $E_F$ exhibits two-dimensional character with flat band along H-K and large dispersive bands along K-Γ and M-Γ. An indirect band gap $E_g$, 0.39 eV, shows up between Γ and M points, indicating that $NaCrTe_2$ is a semiconductor under the A-type AFM state. The metallic conductivity behavior shown in Fig. 2a may come from a little deficiency of Na. The electronic band structure and PDOS of Ising-FM $NaCrTe_2$ are plotted in Fig. 4b. Around $E_F$, the conduction bands along Γ-M and Γ-K are split wider, lowering the conduction band minimum (CBM) and slightly lifting the valence band maximum (VBM). Although the positions of CBM and VBM are unchanged, the $E_g$ becomes narrower to be 0.11 eV. More interestingly, the valence band (VB) and conduction band (CB) are spin-split because the VBM and CBM possess the same spin direction, respectively. It is insulating in spin-down channel with spin-flip gaps $\Delta_1 = 0.4$ eV in VB and $\Delta_3 = 1.8$ eV in CB, and $\Delta_2 = E_g = 0.11$ eV in the spin-up channel.

Thus the NaCrTe$_2$ changes into a FM half-semiconductor under high-$H$ [26],[27],[28]. In Fig. S10, we plotted the electronic band structure and PDOS of NaCrTe$_2$ under Ising-FM and XY-FM states using PBE + U + SOC method. The band topology and DOS values near $E_F$ are identical. Therefore, at $H > H_{sf2} > H_{sf1}$, two FM states exhibit identical low-$R$ state, see Fig. S9. The schematic changes of electronic band structure under $H$ are shown in Fig. 4c and d. Similar $H$-dependent conductivity is proposed in CrI$_3$, in which the change of spin-alignment in turn alters the direct bandgap (0.91 eV) into indirect bandgap (0.96 eV) and thus induces an large MR and AMR ratio [29].

In addition, we observe two different transition trends from high-$R$ to low-$R$ states of NaCrTe$_2$ in Fig. 4e. For $H // c$, the high-$R$ rapidly decreases to low-$R$ value as $H$ is close to $H_{sf1}$. While for $H // ab$, the high-$R$ state slowly approaches low-$R$ state as increasing $H$ to $H_{sf2}$. Such kind of decrement leads to that the difference in $R$ exhibits the maximal value at $H_{sf1}$, determining the largest magnitude of AMR ratio. We calculated the MAE between two FM states, namely rotating the spin direction rotates from $c$-axis ($\varphi = 0°$, Ising-type) to $a$-axis ($\varphi = 90°$, XY-type). In Fig. 4f, it is found that the value of MAE $\xi_{MAE} = E_{100} - E_{001}$ is 1017 μeV per Cr$^{3+}$, which theoretically explains the reason why flipping the high-$R$ state to Ising-type low-$R$ state is easier. It is noted that the MAE is considerably higher than those of known CrI$_3$ (650 μeV) [30,31], Cr$_2$Ge$_2$Te$_6$ (63 μeV) [32,33] and VI$_3$ (290 μeV) [34].

In a typical La$_{0.7}$Ca$_{0.3}$MnO$_3$ single crystal, there are no AMR signals either in pure AFM insulating phase or $H$-induced FM metallic phase. The AMR signal only emerges at the regime of MIT transition, and reaches a peak value at $T$ close to $T_{MIT}$. It can be explained by the competing scenario between AFM insulator and FM metal, induced by Jahn-Teller distortion and double-exchange interaction, respectively. However, the large AMR ratio observed here is significantly different from other known CMR materials. In our work, the AMR ratio shows peak value at 10 K, far away from the $T_N$ or $T_c^{'}$. It recalls the AMR effect in Sr$_2$IrO$_4$, where the largest AMR ratio is found at the base temperature and around $H_{sf}$, where the spin-tuned conductivity drives high-$R$ state to low-$R$ state, leading to a negative MR and sizable AMR ratio [14,15].

## IV. Conclusion

Hence, the spin-flipping-driven magneto-transports in NaCrTe$_2$ can be understood by below two aspects. Firstly, the reduction of $E_g$ from AFM semiconductor to FM half-semiconductor is as large as 280 meV, which is a remarkably rare case. It reflects the strong coupling among spin, lattice and conductivity in NaCrTe$_2$. This kind of change enhances the carrier concentration $n$ and reduces $R$, which is the intrinsic origin of the large negative MR. Similarly, spin-flip transitions are observed in a layered A-type AFM CrSBr under $H$, and the strong coupling between spin and charge leads to negative MR signals of ~40% [35]. Secondly, the MAE between Ising-FM and XY-FM plays the most decisive role in producing AMR signal because it directly determines the drop-off rate from high-$R$ state to low-$R$ state. Assuming that the MAE is very small, the decrement of $R$ to either Ising-like FM or XY-like FM low-R state would be nearly synchronous, and the AMR ratio should be rather weak. Therefore, the reduced magnitude and the different descending in $R$ conjointly determine the large AMR effect. Our work deepens the understanding of origin for spin-related transport property in layered antiferromagnet. More AFM-based materials exhibiting considerable magneto-transport effect are highly expected.

**Acknowledgements**

We thank Dr. Gang Xu and Dr. Zengwei Zhu for measuring the transport property at high magnetic field. This work was supported by the National Key Research and Development Program of China (No. 2017YFA0304700, 2016YFA0300600, 2018YFE0202601 and 2016YFA0300504), National Natural Science Foundation of China (No. 51922105, 51772322, 11774423 and 11822412), Strategic Priority Research Program and Key Research Program of Frontier Sciences of the Chinese Academy of Sciences (No. QYZDJ-SSW-SLH013) and Beijing Natural Science Foundation (Grant No. Z200005).

‡These authors contributed equally to this work.

Email: jgguo@iphy.ac.cn; gaoguoying@ysu.edu.cn; xlchen@iphy.ac.cn

Table 1. Optimized lattice constants (Å) of NaCrTe$_2$ with A-type AFM and Ising-type FM state, total energy of 2×√3 supercell (GGA+U+SOC, eV), bandgap $E_g$ (meV), magnetic coupling constants $J_1$, $J_2$ and $J_3$ (meV), magnetic anisotropy energy (MAE, μeV per Cr$^{3+}$), Néel temperature ($T_N$) and Curie temperature ($T_c^{'}$).

| NaCrTe$_2$ | $a$ | $c$ | $E_{total}$ | $E_g$ | $J_1$ | $J_2$ | $J_3$ | MAE | $T_N$ | $T_c^{'}$ |
|---|---|---|---|---|---|---|---|---|---|---|
| A-type AFM | 4.0053 | 7.4458 | -140.8182 | 0.39 | 2.98 | 1.86 | -0.08 | --- | 110 | --- |
| Ising FM | 4.1412 | 7.3941 | -140.8123 | 0.11 | --- | --- | --- | 1017 | --- | 113 |

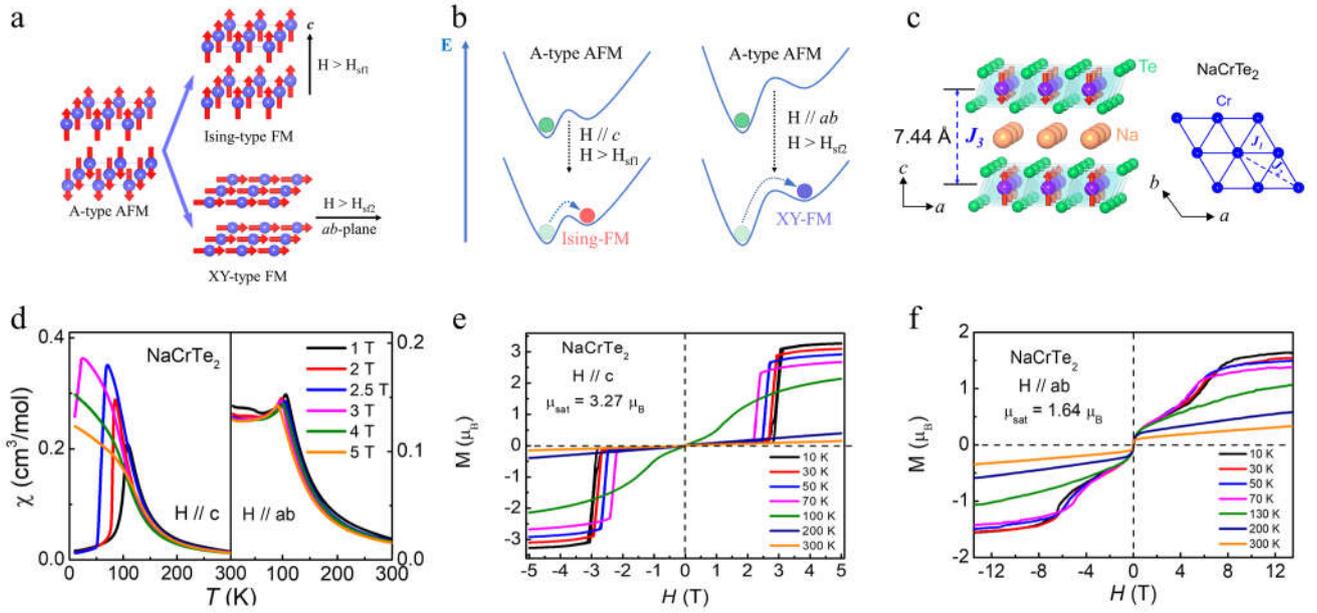

**Figure 1.** Spin-flipped transition of A-type antiferromagnet NaCrTe$_2$. (a) Spin-flip transition in an A-type layered antiferromagnet under *H*. By applying large enough *H* along easy-axis and hard-axis, two kinds of FM states can be obtained. (b) Relative total energy of A-type AFM, Ising-type FM and XY-type FM. (c) Crystal structure of NaCrTe$_2$. Blue ball is Cr, green ball Te and orange ball Na. Interlayer spacing is 7.44 Å. Right panel shows a single Cr layer. $J_1$, $J_2$ and $J_3$ represent the nearest, the next-nearest neighbor and interlayer magnetic exchange constants, respectively. (d) Temperature dependence of magnetic susceptibility ($\chi$) of NaCrTe$_2$ under *H // c* and H//*ab*. (e), (f) Isothermal M-H curves of NaCrTe$_2$ under *H*//*c* and *H*//*ab*.

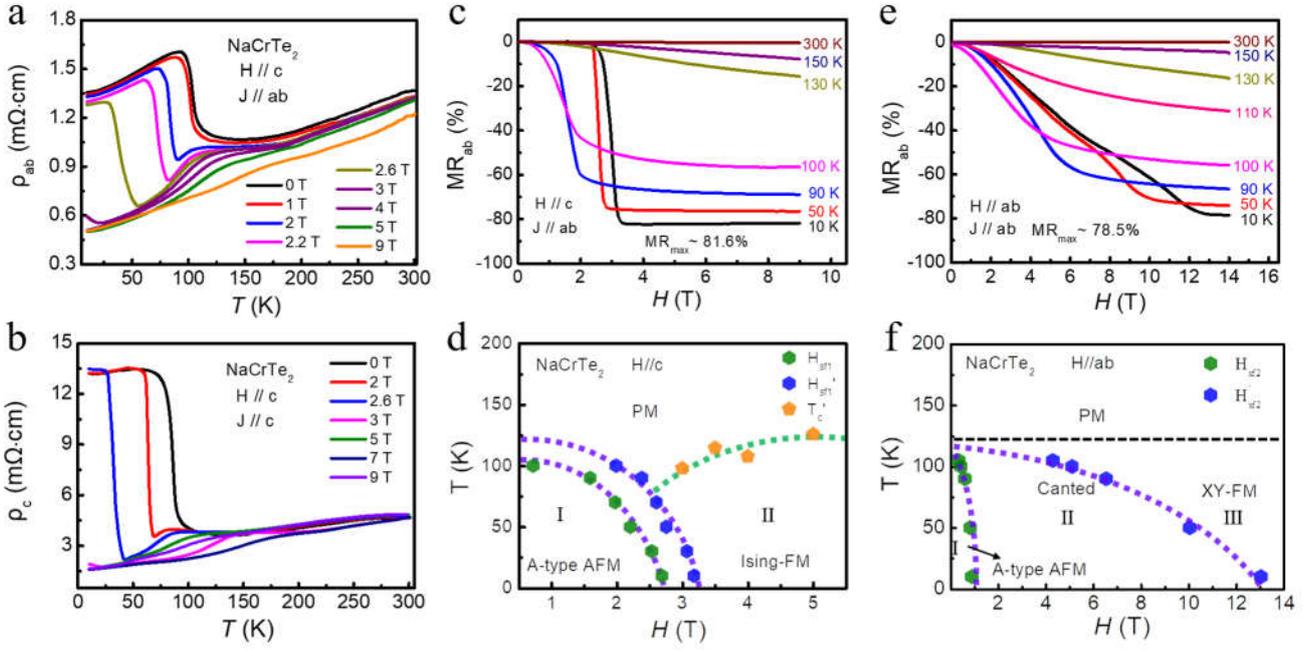

**Figure 2.** Electrical transport and magneto-resistance (MR) of NaCrTe$_2$ single crystal. (a), (b) T-dependent electrical resistivity $\rho_{ab}$ and $\rho_c$ under $H//c$. With increasing $H$, the jump due to PM-to-AFM transition is gradually suppressed. (c), (d) Isothermal MR$_{ab}$ ($J//ab$) of NaCrTe$_2$ at under $H//c$ and $H//ab$. (e), (f) Magnetic phase diagrams in NaCrTe$_2$ under $H//c$ and $H//ab$. The data are taken from the onset-$H$ ($H_{sf1}$, green) of $R$-drop and onset-$H$ ($H'_{sf1}$, blue) of low-$R$ state. $H$-induced FM transition temperature $T_c'$ (orange) are determined from the derivative of M-T curves at high-$H$.

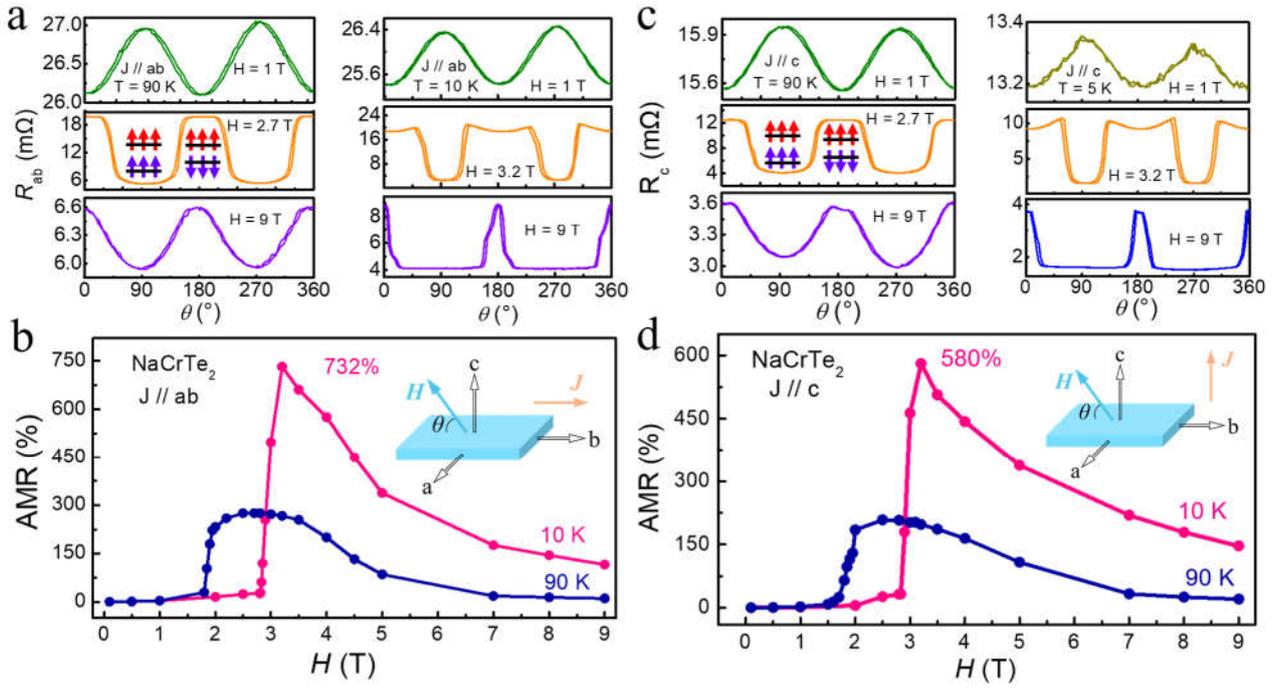

**Figure 3.** Anisotropic magneto-resistances of NaCrTe$_2$ single crystal. (a) $R_{ab}$ of NaCrTe$_2$ single crystal (sample #1) under various $H$ at 90 K and 10 K. (b) $R_c$ of NaCrTe$_2$ single crystal (sample #2) under various $H$ at 90 K and 10 K. The data at selected H are presented for clarity. Insets arrows show schematic spin alignments under given $H$ and angle. (c), (d) Giant AMR ratios are observed in $R_{ab}$ and $R_c$. Insets are the schematic geometry of the relative orientation between current ($J$), magnetic-field ($H$), and crystal axis of slab sample. All measurements were carried out by rotating the angle $\theta$ from 0° to 360° under a specific $H$.

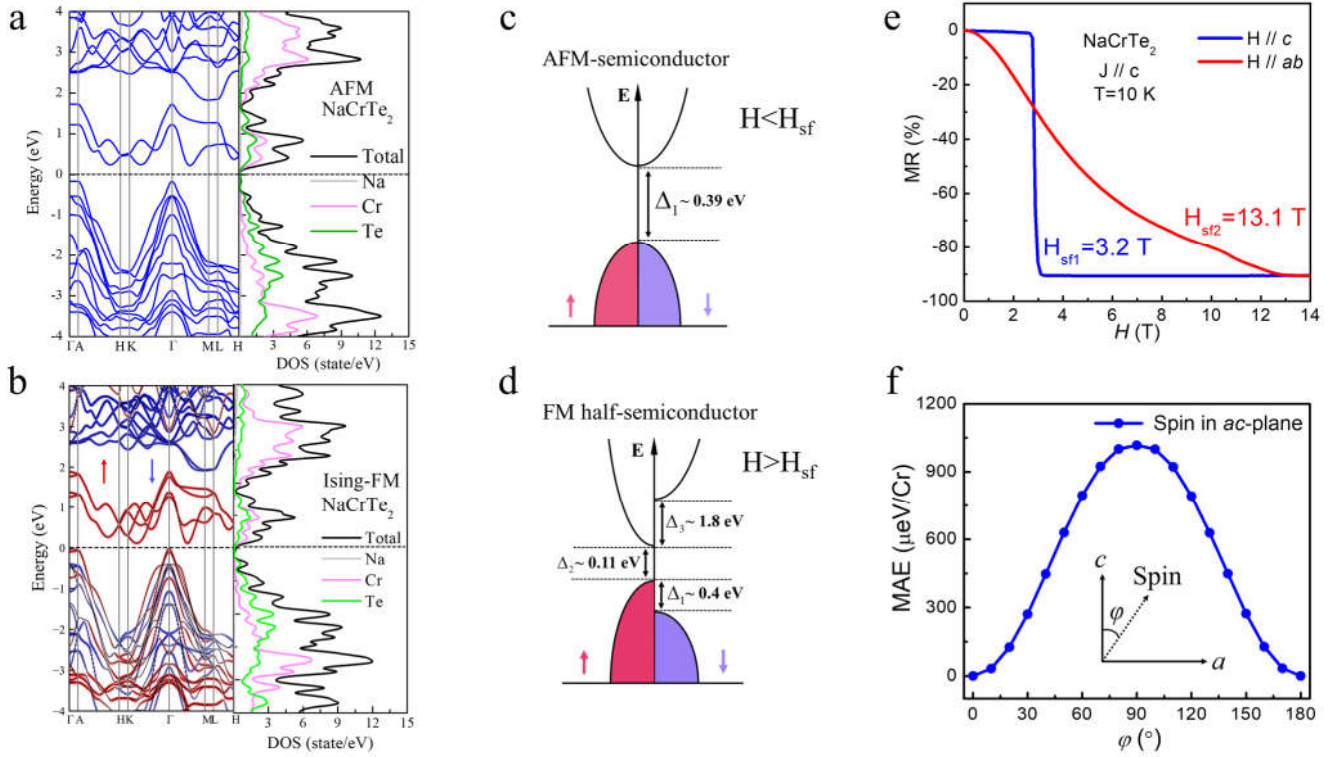

**Figure 4.** Electronic structure and origin of giant AMR. (a), (b) Band structure and PDOS of the A-type AFM and Ising-FM $NaCrTe_2$ calculated by HSE06+U+SOC. (c,d) Schematic electronic structure of A-type AFM semiconductor and FM half-semiconductor under different $H$. (e) MR under $H//c$ and $H//ab$ at 10 K measured from 0 to 14 T. (f) MAE per $Cr^{3+}$ of FM-$NaCrTe_2$. In the inset, the $\varphi=0°$ and $90°$ represent the Ising-type and XY-type FM configurations, respectively.